\documentstyle[11pt,newpasp,twoside,epsf]{article}
\def\edcomment#1{\iffalse\marginpar{\raggedright\sl#1\/}\else\relax\fi}
\reversemarginpar

\begin{document}
\title{The luminosity calibration of the $uvby-\beta$ photometry}
 \author{C. Jordi, X. Luri, E. Masana, J. Torra, F. Figueras}
\affil{Dept. Astronomia i Meteorologia, Univ. de Barcelona,
Avda. Diagonal 647, E-08028 Barcelona, Spain}
\author{A. Domingo}
\affil{LAEFF, INTA, Apartado 50727, E-28080 Madrid, Spain}
\author{A.E. G\'omez }
\affil{Obs. de  Paris-Meudon, DASGAL, F-92195 Meudon Cedex, France}
\author{M.O. Mennessier}
\affil{Univ. Montpellier II, GRAAL,  F-34095
Montpellier Cedex 5, France}

\begin{abstract}
The ESA HIPPARCOS satellite has provided astrometry
of unprecedented accuracy, allowing to reassess, improve and refine
the pre-HIPPARCOS luminosity calibrations. 
We review the "classical" absolute magnitude calibrations with the
Str\"omgren-Crawford in\-ter\-me\-dia\-te-band photometric system.
A small zero point correction of about 2-4\% seems necessary, as well as to 
refine the dependences on metallicity and projected rotational velocity.
The need of a rigorous statistical treatment of the extremely precise HIPPARCOS data 
to derive definite dependences of the luminosity on physical 
stellar parameters is emphasized.

\end{abstract}

\section{The pre-HIPPARCOS luminosity calibrations}

The $uvby-\beta$ photometric system is well suited to derive stellar
physical parameters and in particular the luminosity, through calibrations
accounting for the dependence on $T_{\rm eff}$ and evolution. Further
dependences on metallicity and projected rotational velocity were considered
by several authors but the results were not conclusive enough.

The most widely used calibrations are fully empirical (Crawford 1975, 1978, 
1979; Str\"omgren 1966; Olsen 1984 and Balona \& Shobbrook 1984 among others).
Open clusters and young stellar associations were used to
derive the shape of the ZAMS and the dependence on the evolution, while
the zero point was directly or indirectly fixed through the few trigonometric 
parallaxes precise enough available at the epoch.
A different approach was taken by Nissen \& Schuster (1991), who included
theoretical stellar evolutionary models 
to derive metallicity dependences of the ZAMS. 
The unavoidable need to transform from theoretical to observed 
stellar quantities adds a source of uncertainty, however.

Figure 1 shows how the ground-based parallaxes used to fix the zero point
of the luminosity 
(17 F-type stars closer than 11 pc, $\sigma_{\pi}/\pi < 10.5$\%) 
compare with the HIPPARCOS parallaxes ($\sigma_{\pi} < 1$ mas). 
HIPPARCOS parallaxes are smaller by $5\pm3$ mas, i.e. the stars are 
$0.09\pm0.06$ mag brighter than previously considered. Thus, the photometric
distances derived with the pre-HIPPARCOS calibrations are underestimated 
roughly by about 4\%.

\begin{figure}[h]
\plotone{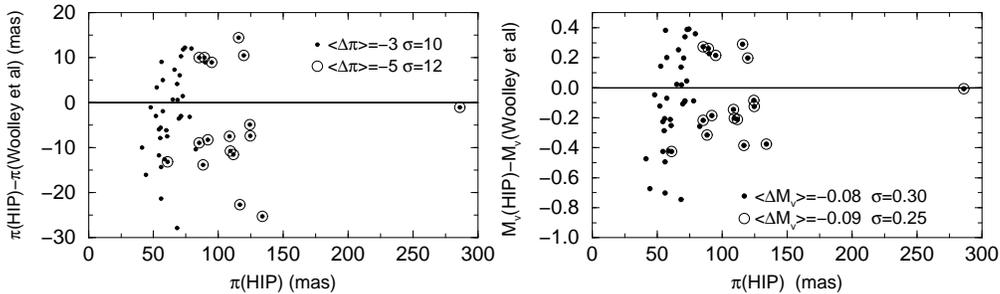}
\caption{{\em Left}: Comparison of Woolley et al. (1970) and HIPPARCOS parallaxes 
for the stars
used to establish the luminosity calibration of F-type stars
(circled points represent those stars used to fix the zero point).
{\em Right}: Differences of $M_v$ obtained using
HIPPARCOS and Woolley et al. trigonometric parallaxes}
\end{figure}

\section {HIPPARCOS based luminosity calibrations}

HIPPARCOS did not only improve the existent trigonometric parallaxes. It
considerably enlarged the sample of stars with precise parallaxes, provided 
accurate proper motions and new data on duplicity and variability. All this
information yields larger and cleaner samples than before. However, a big
amount of precise data is not enough. 
To exploit the HIPPARCOS data to its full extent, a careful evaluation of 
the sample observational selection effects is needed, as well as taking into
account the involved observational errors. Furthermore,
statistically robust treatments accounting for these items
are necessary.

We describe in the following sections the calibrations done in this direction.

\subsection{F-type stars}

A sample of unreddened stars with $2.59 < \beta < 2.72 $ ($\approx$ F0-G2) 
was built from the HIPPARCOS catalogue and Hauck \& Mermilliod (1998, HM)
photometric data. The sample was cleaned of luminosity classes
I \& II, peculiar, variable, binaries and emission stars according to 
the HIPPARCOS flags, SP information, "type object" in SIMBAD and flags
in HM. The sample contains $\sim$700 stars with $\sigma_{\pi}/\pi<5$\% and 
$\sim$2500 stars with $\sigma_{\pi}/\pi<15$\% (about two orders of
magnitude more stars than those available to pre-HIPPARCOS calibrations). 

Figure 2 (left) compares the photometric absolute magnitude with 
the HIPPARCOS data for the first subsample.
Crawford's calibration shows a small zero point difference, as expected
from the parallax differences of the zero point stars (see Fig. 1). A slight 
trend
on metallicity is observed when working with the subsample with 
$\sigma_{\pi}/\pi<15$\%. A similar comparison using Nissen \& Schuster (1991) 
calibration reveals that it is more appropriate 
for the metal poor stars than Crawford's one, which performs better
for non-metal poor stars.

\begin{figure}[h]
\plottwo{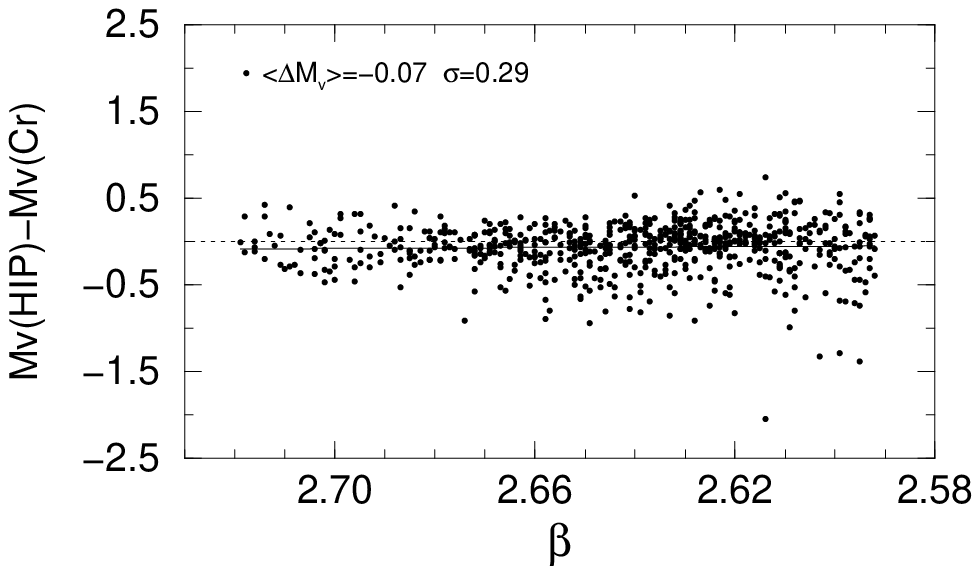}{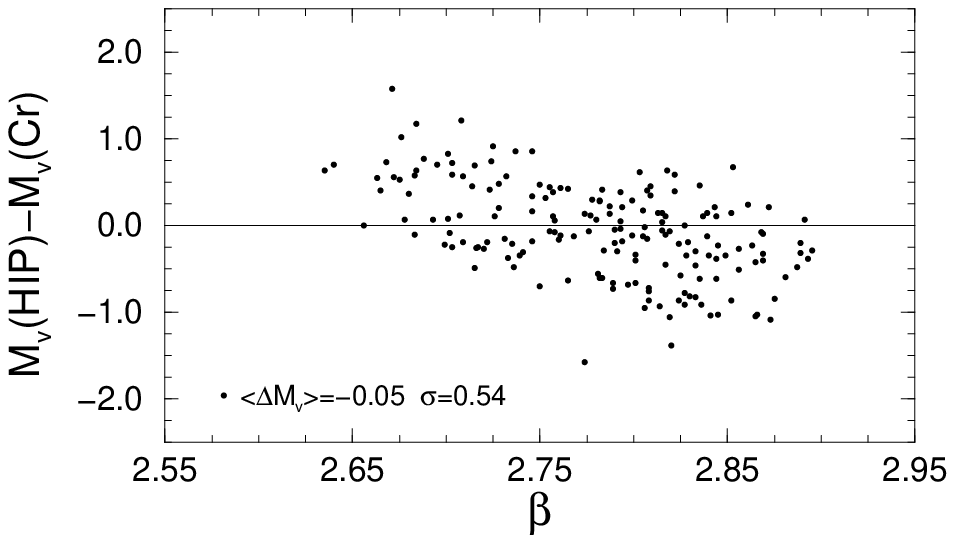}
\caption{Differences of $M_v$ in the sense HIPPARCOS$-$Crawford (1975) 
as a function of effective temperature for the 
F-type stars with $\sigma_{\pi}/\pi < 5$\% ({\em left}) and 
HIPPARCOS$-$Crawford (1978) as a function of luminosity for the
B-type stars with $\sigma_{\pi}/\pi < 10$\% ({\em right})}
\end{figure}

Biases due to the observational selection effects and errors were tested using
MonteCarlo simulations. A sample of stars of a given $\overline{M_v}$ with
spatial distribution typical of the disc
(assuming $\sigma_{M_v}=0.3$ mag, typical HIPPARCOS errors for
parallax and appropriate limits on apparent magnitude) was generated.
When the $M_v$ is derived by just using $\pi_{\rm obs}$ and the sample
is limited to contain stars with $\sigma_{\pi}/\pi<5\%$, we found that 
$\overline{M_v({\rm true})-M_v(\pi_{\rm obs})} =-0.01$ if $\overline{M_v}=2.7$ 
mag ($\sim$ F0) and
$\overline{M_v({\rm true})-M_v(\pi_{\rm obs})} =+0.06$ if $\overline{M_v}=4.7$ mag
($\sim$ G2).

With those biases in mind, we attempted to derive a new calibration by a
simple least square fit. The results point out to $\sim 30\Delta\beta$
more likely than the $20\Delta\beta$ adopted by Crawford (1975) as evolutionary
term, to a small correction for metallicity ($\sim 0.2 [Fe/H]$) and to a 
slightly different $M_v({\rm ZAMS)}$ relation ($ <0.2$ mag) . The dispersion of the residuals is of 0.25
mag only slightly better than the 0.29 mag (Fig. 2, left) of Crawford's calibration
and larger than expected from the $\sigma_{\pi}/\pi$ value,
reflecting the contribution of the photometric errors, unsolved
binarity, differences in helium composition, cromospheric activity and so on.
The continuity with the 
A-type stars (see next subsection) has not been tested yet. A more conclusive 
calibration including the observational errors will be the subject
of a forthcoming paper.

\subsection{Late A-type stars}

Domingo \& Figueras (1999) compared photometric and HIPPARCOS parallaxes
for a sample of normal and metallic A-type stars and they concluded that
Crawford (1979) calibration
is better than Guthrie (1987) calibration for normal A-type stars,
while
Guthrie's calibration is better than Crawford's one
for Am stars. 
In both cases, according to HIPPARCOS, the stars
are brighter than predicted by the old calibrations, as for the F-type stars. 

A new calibration (including normal and Am stars) was derived by these 
authors using the weighted least square and the BCES methods 
and working with a sample of stars 
closer than 100 pc (including individual Lutz \& Kelker correction).
The precision of their calibration is of 0.23 mag. 
The dependences on evolution, blanketing and rotational velocity are
smaller than the ones derived in the pre-HIPPARCOS calibrations. 

\subsection{B-type stars}

As for the F-type stars, a sample of B-type stars was built with similar
selection criteria. It contains $\sim$1500 stars ($\sim$ 20 stars with
$\sigma_{\pi}/\pi < 5$\% and $\sim$200 stars with $\sigma_{\pi}/\pi < 10$\%). 
Photometric and HIPPARCOS based absolute magnitudes are compared in Fig. 2 
(right) for Crawford (1978) calibration.  Similar
simulations that for the F-type stars were performed and we obtained that 
$\overline{M_v({\rm true})-M_v(\pi_{\rm obs})}=-0.10$ mag at $\beta=2.65$ and 
$\overline{M_v({\rm true})-M_v(\pi_{\rm obs})}=-0.05$ mag at $\beta=2.90$ for a
sample limited to 
$\sigma_{\pi}/\pi < 10$\%. Thus, the biases
are smaller than the actual differences. When the comparison is performed with
Balona \& Shobbrook (1984) calibration, a similar trend also appears, although less
pronounced.

Due to the scarcity of early B-type stars in the near vicinity of the Sun, 
the calibration should be approached by using
open clusters or by using
a statistical method capable of dealing with the whole sample, accounting for its biases
and the observational errors.
The maximum likelihood method (LM) by Luri et al. (1996) was adapted
to this case. We parameterized the absolute magnitude
as a function of the $\beta$ and $[u-b]$ colours and the results are
quite promising, although the influence of the adopted interstellar absorption
model must be checked.
Luminosity seems to depend only on $\beta$ for 
the B0-B3 subsample, while an additional  dependence on $[u-b]$ exists
in the B4-B9 subsample (Crawford (1978) did not considered an evolutionary
term when $c_o > 0.9$ and this could be the cause of the large discrepancy
at the B8-B9 range in Fig. 2 right). Dependences on projected rotational velocity
are not clear at this stage of the calibration.
Again, a more definite calibration using LM and its comparison with 
the classical approach is the subject of our future work,
although we are limited by
the small number of projected rotational velocities available.

\end{document}